\documentclass[nofootinbib,amsmath,amssymb,amsfonts,aps,prd,eqsecnum,superscriptaddress]{revtex4-1}

\pdfoutput=1
\usepackage{hyperref}
\usepackage{amsmath}
\usepackage{amssymb}
\usepackage{mathrsfs}
\usepackage{xcolor,graphicx}
\usepackage[caption=false]{subfig}
\usepackage[toc,page]{appendix}
\usepackage[section]{placeins}
\usepackage{breqn}
\usepackage[T1]{fontenc}

\makeatletter
\let\cat@comma@active\@empty
\makeatother

\begin{document}

\title{Exploration of Hamiltonian formulations of baroclinic hydrodynamics}

\author{John Ryan Westernacher-Schneider}
\email{jwestern@email.arizona.edu}
\affiliation{Department of Astronomy/Steward Observatory, 
  The University of Arizona, \\
  933 N. Cherry Ave, Tucson, AZ 85721, USA}

  


\begin{abstract}
A neutron star in a compact binary is expected to be well-approximated by a barotropic flow during the inspiral phase. During the merger phase, where tidal disruption and shock-heating occur, a baroclinic description is needed instead. In the barotropic case, a Hamiltonian formulation potentially offers unique benefits for numerical relativity simulations of the inspiral phase, including highly accurate conservation of circulation and superconvergence of the fluid variables, and is actively being explored. In this work, we investigate the viability of a Hamiltonian formulation in the baroclinic case. At odds with the barotropic case, this formulation is non-conservative, yet it can be treated well with approximate Riemann solver algorithms since the non-conservative terms vanish across genuinely nonlinear fields. Nonetheless, using numerical 1-dimensional shock tube tests we find that the weak solutions of the Hamiltonian system differ from the standard ones obtained by enforcing conservation of rest mass density, momentum density, and energy density across discontinuities. We also show that barotropic Hamiltonian formulations can admit shockwaves at fluid-vacuum interfaces, which may be related to the unstable behavior of stellar surfaces observed in past numerical tests. In light of the unphysical weak solutions, we expect that in future implementations of the Hamiltonian formulation of hydrodynamics in numerical relativity it will be necessary to use an explicitly barotropic formulation during the inspiral phase, and then switch to a robust baroclinic formulation prior to merger. 


\end{abstract}

\maketitle

\section{Introduction}\label{sec:intro}
Gravitational wave astronomy is upon us~\cite{aasi2015advanced, acernese2014advanced, abbott2016observation, abbott2017gw170817, abbott2019gwtc}. Today's gravitational wave data has more uncertainty~\cite{abbott2017effects, haster2016inference} than theoretical models of the signals coming from perturbation theory, phenomenology, or numerical relativity~\cite{buonanno1999effective, husa2016frequency, boyle2019sxs, dietrich2018core, haas2016simulations, kiuchi2017sub}, therefore those models are used to inform interpretations of the data~\cite{abbott2016improved, abbott2019properties}. However, with future third-generation detectors~\cite{abbott2017exploring, reitze2019cosmic} or current detectors at design-sensitivity, the relationship between signal models and data may invert~\cite{samajdar2018waveform, samajdar2019waveform, brown2020data}; data will inform the models. This inversion will take place unless substantial advancements are made in the accuracy of theoretical models, including that of numerical relativity simulations. Advancements are being pursued in perturbative calculations~\cite{goldberger2006effective, galley2009radiation, blanchet2014gravitational, bini2020binary, bini2020sixth, nagar2019efficient, blumlein2020fourth, landry2015tidal, landry2017tidal, pani2015tidal, pani2015tidal2, steinhoff2016dynamical} as well as numerical simulations. Some existing efforts in the latter category involve innovations and optimizations of hardware~\cite{lim2015technological, stevens2019aurora, narasimhamurthy2019sage}, parallel-computing software~\cite{kidder2017spectre, palenzuela2018simflowny, zhang2019amrex,fernando2018massively,neilsen2019dendro}, and numerical methods~\cite{radice2013beyond, miller2016operator, bugner2016solving, bernuzzi2016gravitational, felker2018fourth, foucart2019smooth, most2019beyond, fambri2018ader, fambri2020discontinuous, koppel2018towards, ruchlin2018senr, ruchlin2018senrcode}.

Another strategy is to innovate at the level of the physics formulation, possibly combined with the use of unconventional or innovative numerical methods. The numerical application of a Hamiltonian formulation of fluid dynamics falls into this category. The formulation goes back many years~\cite{synge1937relativistic, lichnerowicz1941invariant}, has been expounded upon recently~\cite{markakis2014hamiltonian, markakis2017conservation}, and proofs-of-principle have begun in numerical applications~\cite{westernacher2019hamilton}. This work is part of a series of papers exploring the numerical applicability of Hamiltonian formulations of fluid dynamics for relativistic stars, with an eye specifically toward gravitational wave-driven decay of binary systems involving at least one material body. 

The inspiral phase of a binary neutron star or black hole-neutron star system is expected to be well-approximated as barotropic~\cite{rieutord2006introduction, friedman2013rotating}, i.e.~only one scalar fluid variable is independent. An example of a barotropic equation of state is that of a polytrope, $P=\kappa \rho_0^\Gamma$, where $P$ is the fluid pressure, $\rho_0$ is the rest mass density, $\Gamma$ is the adiabatic index, and $\kappa$ is a polytropic parameter related to temperature. Inviscid barotropic flow is known to conserve circulation~\cite{kelvin1869vortex}, even during gravitational-wave driven decay~\cite{friedman2013rotating}. A Hamiltonian formulation of barotropic fluid dynamics, such as that considered in~\cite{westernacher2019hamilton}, has several properties which appear suitable for numerics. First, it is likely that a scheme to conserve circulation with high accuracy can be developed with the help of constraint damping~\cite{brodbeck1999einstein, gundlach2005constraint}. Second, since circulation is conserved, an initially irrotational binary will remain irrotational, and by imposing irrotationality in the Hamiltonian formulation, one obtains a genuinely flux-conservative form of the Euler equation valid in arbitrary spacetimes~\cite{markakis2014hamiltonian, markakis2017conservation}. The absence of source terms in such a form of the Euler equation should enable the development, with relative ease, of a general well-balanced numerical scheme (i.e.~a scheme which preserves general equilibrium configurations to within machine precision). The fluid-vacuum interface was found in~\cite{westernacher2019hamilton} to be delicate in the Hamiltonian formulation, which was solved by hybridizing the Hamiltonian formulation with the more common Valencia scheme~\cite{banyuls1997numerical} in the vicinity of the surface. In Appendix~\ref{app:vac} we point out that shockwaves are (unphysically) permitted at fluid-vacuum interfaces in the Hamiltonian formulation, which may be related to the unstable behavior of stellar surfaces observed in~\cite{westernacher2019hamilton}.

During the tidal disruption and merger phase of a binary system, the fluid is not barotropic. Instead, a baroclinic description is required. An example of such a fluid is that with a gamma-law equation of state $P=\rho_0 \epsilon (\Gamma -1)$, where $\epsilon$ is the specific internal energy, and again where $P$ is the fluid pressure, $\rho_0$ is the rest mass density, $\Gamma$ is the adiabatic index. Fewer numerical benefits are currently apparent for a Hamiltonian formulation of baroclinic fluid dynamics, although see~\cite{markakis2017conservation}. However, for eventual applications in numerical relativity, it is important to understand and anticipate the suitability of Hamiltonian formulations of fluid dynamics in different regimes through simple numerical explorations and proofs-of-principle.

In the current work, we numerically explore the weak solutions in Hamiltonian formulations of baroclinic fluid dynamics using a variety of shock tube tests. Such formulations are non-conservative, therefore the question of weak solutions is a nontrivial one both mathematically~\cite{vol1967spaces, cauret1989discontinuous, dal1995definition} and in terms of numerical methods~\cite{diaz2008sediment, pelanti2008roe, fernandez2008new, munkejord2009musta,morales2009shallow, dumbser2011simple, muller2013well, fernandez2014efficient, diaz2014second, sanchez2016hllc, castro2017well}. However, using recent approximate Riemann solvers for non-conservative systems~\cite{dumbser2016new}, we show that the weak solutions in Hamiltonian formulations are in general unphysical, although rarefaction fans and large discontinuities in rest mass density $\rho_0$ (without discontinuities in specific internal energy $\epsilon$) are well-captured. The cause of this is likely due to the evolution variables being unsuitable for the description of shockwaves~\cite{dumbserprivate,toro2013riemann} (the conserved quantities across shockwaves should be rest mass, momentum, and energy), although a rigorous argument to this effect is not forthcoming since the Hamiltonian formulation is non-conservative. The main conclusion of this work is that, in future applications of Hamiltonian formulations to binary simulations, it is advisable to use an explicitly barotropic formulation during the inspiral phase, and then switch to a more robust baroclinic formulation (such as Valencia~\cite{banyuls1997numerical}) at some time prior to merger.

In Sec.~\ref{sec:forms} we develop several baroclinic Hamiltonian formulations. Settling on one formulation, in Sec.~\ref{sec:weak} we discuss the question of weak solutions and appropriate numerical schemes. In Sec.~\ref{sec:results} we present numerical results for several shock tube tests in flat spacetime in 1+1 dimensions. We conclude in Sec.~\ref{sec:conc}, and discuss the vacuum Riemann problem in Appendix~\ref{app:vac}.

Throughout, we use the mostly-positive metric signature $(-,+,+,...)$. Spacetime indices are denoted with letters at the beginning of the alphabet $\lbrace a,b,c... \rbrace$. Spatial indices are denoted with letters beginning in the middle of the alphabet $\lbrace i,j,k... \rbrace$. We use units in which $G=c=1$, and the Boltzmann constant $k_b$ will appear explicitly.

%
%

\section{Baroclinic Formulations} \label{sec:forms}
%
%
In this section we develop several baroclinic Hamiltonian formulations of fluid dynamics. We will settle on one choice when presenting numerical results.

We begin with a perfect fluid with a relativistic gamma-law equation of state, $P= \rho_0 \epsilon (\Gamma -1)$, where $\rho_0$ is the rest mass density, $\epsilon$ is the specific internal energy, and $\Gamma$ is the adiabatic index. The first equation we employ is the continuity equation
\begin{eqnarray}
0 = \nabla_a \left( \rho_0 u^a \right),
\end{eqnarray}
which expresses the local conservation of rest mass, and thus is valid at low energies. Since this is a total divergence, in curved spacetime we obtain
\begin{eqnarray}
\partial_a \left( \sqrt{-g} \rho_0 u^a \right) = 0. \label{eq:rm_totdiv}
\end{eqnarray}
Thus by choosing a densitized variable $\tilde{D} \equiv \sqrt{-g} \rho_0 u^t$, we avoid geometric source terms in this equation. Introduce a 3+1 split of spacetime~\cite{Alcubierre:2008} via
\begin{eqnarray}
ds^2 = (-\alpha^2 + \gamma_{ij}\beta^i \beta^j) dt^2 + 2\gamma_{ij} \beta^i dt dx^j + \gamma_{ij} dx^i dx^j,
\end{eqnarray}
and note the following relations: $\alpha u^t = W = 1/\sqrt{1-\gamma_{ij}v^i v^j}$ is the Lorentz factor, $v^i$ is the fluid 3-velocity measured by normal observers, $u^i/u^t = 
\alpha v^i - \beta^i$ is the advective velocity, and the metric determinant factors via $\sqrt{-g} = \alpha \sqrt{\gamma}$ where $\gamma$ is the spatial metric determinant.

With this infrastructure we obtain the 3+1 form of Eq.~\eqref{eq:rm_totdiv},
\begin{eqnarray}
0 = \partial_t \tilde{D} + \partial_i \left( \alpha \tilde{D} \left(v^i - \beta^i/\alpha \right) \right). \label{eq:Deqn}
\end{eqnarray}

In the barotropic case~\cite{westernacher2019hamilton}, the canonical form of the Euler equation can be written abstractly as $\boldsymbol{u}\cdot \boldsymbol{dp}=0$, where $\boldsymbol{u}$ is the fluid 4-velocity and $\boldsymbol{dp}$ is the exterior derivative of the canonical momentum $p_a = h u_a$ (also known as the canonical vorticity 2-form), where $h$ is the specific enthalpy. In the baroclinic case~\cite{markakis2017conservation}, the canonical form of the Euler equation has an additional term proportional to the gradient of the specific entropy $S$:
\begin{eqnarray}
\boldsymbol{u}\cdot \boldsymbol{d} \boldsymbol{p} = T \boldsymbol{d} S. \label{eq:canon_baroclinic}
\end{eqnarray}
%
The canonical momentum has the same form in the baroclinic case, $p_a = h u_a$, where $h$ is the specific enthalpy. The equation governing $p_i$ will be the spatial part of Eq.~\eqref{eq:canon_baroclinic} in the given chart, and the evolved variable is $p_i$. We obtain
\begin{eqnarray}
0 = \partial_t p_i - \partial_i p_t + \left(\alpha v^j - \beta^j \right) \omega_{ji} - \frac{\alpha T}{W} \partial_i S. \label{eq:canon2_clin}
\end{eqnarray}

With one additional variable compared with the barotropic case, we require one additional equation of motion to close the system. One possible choice is to evolve the energy equation,
\begin{eqnarray}
0 = \nabla_a T^{at}.
\end{eqnarray}
Doing so will result in geometric source terms since this is not a pure covariant divergence of the form $\nabla_a A^a$ for some vector $A^a$. In our numerical tests we found it is possible to evolve the resulting system, although it is rather unstable with discontinuous initial data and equilibrium stars, so we do not pursue it further.

A different choice for the third equation comes from Eq.~\eqref{eq:canon_baroclinic}, obtained by projecting it onto $u^a$:
\begin{eqnarray}
0 = u^a \nabla_a S. \label{eq:Seqn}
\end{eqnarray}
In a chart $\lbrace t, x^i \rbrace$, this is simply the advection of entropy
\begin{eqnarray}
0 &=& \partial_t S + \frac{u^i}{u^t} \partial_i S \nonumber\\
&=& \partial_t S + \left(\alpha v^i - \beta^i \right) \partial_i S \label{eq:Sform}\\
\Rightarrow 0 &=& \partial_t e^S + \left(\alpha v^i - \beta^i \right) \partial_i e^S, \label{eq:ExpSform}
\end{eqnarray}
where the last line anticipates that the entropy is a logarithm.

Numerically we found greatest stability with yet another choice, which is to combine Eq.~\eqref{eq:ExpSform} with the rest mass conservation Eq.~\eqref{eq:Deqn} to obtain a flux-conservative equation,
\begin{eqnarray}
0 &=& \partial_t \left( \tilde{D} e^S \right) + \partial_i \left( \alpha \tilde{D} e^S \left(v^i - \beta^i/\alpha\right) \right). \label{eq:ConExpSform}
\end{eqnarray}
This comes from the covariant conservation of the entropy current $\nabla_a (\rho_0 S u^a) = 0$, except we have chosen to exponentiate the specific entropy (which is also a valid form of the equation).

In order to relate $S$ to the other variables $\lbrace \rho_0, \epsilon \rbrace$, we use the ideal gas equation of state in its other form, $P \propto n k_b T$, where $n$ is the number density of particles and $k_b$ is the Boltzmann constant. Note we use a proportionality sign in order to also accommodate a photon gas, since a special proportionality constant appears in that case involving the Riemann zeta function~\cite{schwabl2006statistical}. Then we determine the specific entropy $S$ via the first and second laws of thermodynamics,
\begin{eqnarray}
T(\rho_0,\epsilon) dS &=& d\epsilon + P d\left( \frac{1}{\rho_0} \right) \nonumber\\
&=& d\epsilon - \frac{P}{\rho_0^2} d\rho_0.
\end{eqnarray}
With $S=S(\rho_0,\epsilon)$ we have the system
\begin{eqnarray}
\frac{\partial S}{\partial \epsilon} &=& \frac{1}{T(\rho_0,\epsilon)} \label{eq:ent1}\\
\frac{\partial S}{\partial \rho_0} &=& -\frac{P}{\rho_0^2}\frac{1}{T(\rho_0,\epsilon)}. \label{eq:ent2}
\end{eqnarray}
We next plug in $P = \rho_0 \epsilon (\Gamma -1)$ and $\epsilon = a T$ for some proportionality constant $a$, which for a material fluid is $a= (k_b/m)(\Gamma -1)^{-1}$ and $m$ is the (average) particle mass composing the fluid. Then Eq.~\eqref{eq:ent1} yields
\begin{eqnarray}
S = a\ln{\epsilon} + f(\rho_0),
\end{eqnarray}
and the integration function $f(\rho_0)$ is determined by Eq.~\eqref{eq:ent2},
\begin{eqnarray}
\frac{\partial f}{\partial \rho_0} = -a(\Gamma-1) \frac{1}{\rho_0}. 
\end{eqnarray}
This gives
\begin{eqnarray}
S = a \left( \ln{\epsilon} - (\Gamma -1) \ln{\rho_0} \right) + S_0,
\end{eqnarray}
with $S_0$ an arbitrary reference level. When plugged into Eqs.~\eqref{eq:ExpSform} or~\eqref{eq:ConExpSform}, the constants $a$ and $S_0$ drop out. Thus, we can replace $S$ in those equations with
\begin{eqnarray}
\hat{S} &\equiv & \ln{\epsilon} - (\Gamma -1) \ln{\rho_0} \nonumber\\
&=& \ln \left( \epsilon \rho_0^{1-\Gamma} \right),
\end{eqnarray}
where we have set $a=1$ and $S_0=0$.

In the Euler equation~\eqref{eq:canon2_clin}, the entropy term $ (\alpha/W) T \partial_i S$ is $(\alpha/W) (\epsilon/a) \partial_i (a \hat{S})$, thus we have
\begin{eqnarray}
0 = \partial_t p_i - \partial_i p_t + \left(\alpha v^j - \beta^j \right) \omega_{ji} - \frac{\alpha\epsilon}{W} \partial_i \hat{S}. \label{eq:EulerS}
\end{eqnarray}
From here onward we will redefine $S\equiv \hat{S}$ to ease notation.
The form of Eq.~\eqref{eq:EulerS} is appropriate for the formulation which evolves $S$, i.e.~Eq.~\eqref{eq:ExpSform}, since then the non-conservative term is written in terms of derivatives of the evolved variables.

If evolving $e^{S}$ instead, then one should write
\begin{eqnarray}
0 = \partial_t p_i - \partial_i p_t + \left(\alpha v^j - \beta^j \right) \omega_{ji} - \frac{\alpha\epsilon}{W e^{S}} \partial_i e^{S}.
\end{eqnarray}

We focus instead on the formulation using the flux-conservative form of the entropy equation~\eqref{eq:ConExpSform}, which motivates the following rewriting:
\begin{eqnarray}
0 = \partial_t p_i - \partial_i p_t + \left(\alpha v^j - \beta^j \right) \omega_{ji} + \frac{\alpha}{W}\frac{\epsilon}{\tilde{D}} \partial_i \tilde{D} -\frac{\alpha}{W} \frac{\epsilon}{\tilde{D} e^S} \partial_i \left( \tilde{D}e^S \right). \label{eq:EulerConExpS}
\end{eqnarray}
In this approach, the evolution variables are $\lbrace \tilde{D} \! =\! \! \sqrt{|g|} \rho_0 u^t = \sqrt{\gamma} \rho_0 W,\; p_i \! =\! h u_i,\; \tilde{D} e^S = \sqrt{\gamma} \epsilon W/\rho_0^{\Gamma-2} \rbrace$. Optionally one can also evolve the vorticity tensor $\omega_{ij}$ as an independent variable using
\begin{eqnarray}
\partial_t \omega_{ij} + \partial_m \left[ \left(\alpha v^k - \beta^k\right) \left( \omega_{kj} \delta^m_i - \omega_{ki} \delta^m_j \right) \right] + \partial_j \left( \frac{\alpha \epsilon}{W} \right) \partial_i S - \partial_i \left( \frac{\alpha \epsilon}{W} \right) \partial_j S = 0.
\end{eqnarray}
This is flux-conservative in the barotropic case, since $\partial_i S =0$, and constitutes the differential form of Kelvin's circulation theorem. But in the baroclinic case one would have to deal with terms quadratic in spatial derivatives.

\subsubsection{Recovery of primitive variables}

Recovery of the primitive variables from the evolution variables proceeds via iterative rootfinding on a function of the rest mass density. 
To recover the primitive variables, one can write
\begin{eqnarray}
e^{S} \tilde{D}^{\Gamma-1} = \frac{h-1}{\Gamma} \left( \sqrt{\gamma} W \right)^{\Gamma -1},
\end{eqnarray}
and then eliminate $W$ in favor of $p^t$ using $p^t = h u^t = h W/\alpha$. This yields
\begin{eqnarray}
e^{S} \tilde{D}^{\Gamma-1} = \frac{h-1}{\Gamma} \left( \sqrt{\gamma} \frac{\alpha p^t}{h} \right)^{\Gamma -1}. \label{eq:rootfind_clin}
\end{eqnarray}
Finally, eliminate $p^t$ in favor of $h$ using $p_a p^a = -h^2 \rightarrow \alpha p^t = \sqrt{\gamma^{ij} p_i p_j + h^2}$, yielding an equation where the only unknown is $h$. 
Once $h$ is solved for, the specific internal energy $\epsilon$ is recovered via $h = 1 + \epsilon \Gamma$. Then the velocity is found via the canonical momentum $p_i$ and the rest mass density is found via $\tilde{D} = \sqrt{\gamma} \rho_0 W$.
%
%
\subsection{Specialization to Minkowski space in $d+1$ dimensions}

In Minkowski space in Cartesian coordinates, we have $\alpha=1$, $\sqrt{\gamma}=1$, and $\tilde{D} = \rho_0 W \equiv D$. Then Eq.~\eqref{eq:rootfind_clin} combined with $p^t = \sqrt{\delta^{ij} p_i p_j + h^2}$ becomes
\begin{eqnarray}
\Rightarrow \;\;\; \left( e^{S} D^{\Gamma -1} \right)^2 h^{2(\Gamma -1)} &=& \frac{(h-1)^2}{\Gamma^2} \left( h^2 + p_i p^i \right)^{\Gamma -1}.
\end{eqnarray}
Just as in the curved space case Eq.~\eqref{eq:rootfind_clin}, in general this will yield a high-order polynomial with no analytic solution, and will thus require an iterative rootfinder to solve for $h$. However, we will consider $\Gamma = 2$ which yields the following quartic polynomial, the highest order polynomial with analytic solutions:
\begin{eqnarray}
h^4 - 2h^3 + \left[ p_i p^i - \left(2 e^{S} D \right)^2 + 1\right] h^2  - 2 p_i p^i h + p_i p^i = 0.
\end{eqnarray}
We checked the $\Gamma=2$ solution space for a wide variety of physical values for the hydrodynamic variables, and in all cases only one physical root existed (i.e. $h \in \mathbb{R}$ and $h\geq 1$).

%
%

%
%
\section{Weak solutions and numerical schemes} \label{sec:weak}
The formulation we focus on consists of the equations of motion~\eqref{eq:Deqn},~\eqref{eq:EulerConExpS},~\eqref{eq:ConExpSform}, which we collect here:
\begin{eqnarray}
0 &=& \partial_t \tilde{D} + \partial_i \left( \alpha \tilde{D} \left(v^i - \beta^i/\alpha \right) \right), \label{eq:Dcollected}\\
0 &=& \partial_t p_i - \partial_i p_t + \left(\alpha v^j - \beta^j \right) \omega_{ji} + \frac{\alpha}{W}\frac{\epsilon}{\tilde{D}} \partial_i \tilde{D} -\frac{\alpha}{W} \frac{\epsilon}{\tilde{D} e^S} \partial_i \left( \tilde{D}e^S \right), \label{eq:pcollected} \\
0 &=& \partial_t \left( \tilde{D} e^S \right) + \partial_i \left( \alpha \tilde{D} e^S \left(v^i - \beta^i/\alpha\right) \right), \label{eq:DeScollected}
\end{eqnarray}
together with the equation of state $P=\rho_0 \epsilon (\Gamma -1)$. The system has the form
\begin{eqnarray}
0 &=& \partial_t \boldsymbol{Q} + \partial_i \boldsymbol{F} + B(\boldsymbol{Q}) \partial_i \boldsymbol{Q}, \label{eq:nonconsform}
\end{eqnarray}
where $\boldsymbol{Q}$ is the evolution variables collected into a vector, $\boldsymbol{F}$ is a flux, and $B$ is a square matrix encapsulating the non-conservative part of the system. 

The very notion of weak solutions was extended to non-conservative systems through the use of Borel measures~\cite{dal1995definition}\footnote{Earlier work on the subject can be found in e.g.~\cite{vol1967spaces,cauret1989discontinuous}}. In this context, the notion requires a choice of path through solution space which interpolates between states on either side of a discontinuity. There has been a lot of work extending and applying approximate Riemann solvers to non-conservative systems of the form of Eq.~\eqref{eq:nonconsform}, e.g.~\cite{diaz2008sediment, pelanti2008roe, fernandez2008new, munkejord2009musta,morales2009shallow, dumbser2011simple, muller2013well, fernandez2014efficient, diaz2014second, sanchez2016hllc} (see~\cite{castro2017well} and references therein). In this work we use the path-conservative version of the original~\cite{harten1983upstream} Harten-Lax-van Leer (HLL) scheme given explicitly in~\cite{dumbser2016new}, and our conclusions do not change if we instead use the HLLEM scheme (also given explicitly in~\cite{dumbser2016new}).

In Eq.~\eqref{eq:nonconsform}, path-conservative schemes can be shown to reduce to standard conservative schemes when the matrix $B$ is a Jacobian of some flux vector. Usually this is not the case, e.g.~multi-layer shallow water or multi-phase flows~\cite{dumbser2016new}. For the system of Eqs.~(\ref{eq:Dcollected},~\ref{eq:pcollected},~\ref{eq:DeScollected}), in the 1-dimensional case we have $\omega_{ji}=0$, and the matrix $B$ encapsulating the non-conservative terms is
\begin{eqnarray}
B = \left[\begin{matrix}
0 & 0 & 0 \\
\frac{\alpha}{W} \frac{\epsilon}{\tilde{D}} & 0 & -\frac{\alpha}{W}\frac{\epsilon}{\tilde{D} e^S} \\
0 & 0 & 0
\end{matrix} \right]
= \left[\begin{matrix}
0 & 0 & 0 \\
\frac{\alpha}{W^2} \frac{\epsilon}{\sqrt{\gamma} \rho_0} & 0 & -\frac{\alpha}{W^2}\frac{\rho_0^{\Gamma -2}}{\sqrt{\gamma}} \\
0 & 0 & 0
\end{matrix} \right]
\end{eqnarray}
One can easily show that this matrix is not the Jacobian of any flux, since the system of partial differential equations the flux would have to satisfy are inconsistent. Namely, for a flux $\boldsymbol{f}$ such that $B = \partial \boldsymbol{f}/\partial \boldsymbol{Q} = \partial \boldsymbol{f}/\partial \boldsymbol{p} \cdot \partial \boldsymbol{p}/\partial \boldsymbol{Q}$, where $\boldsymbol{p}$ is a vector of primitive variables, it suffices to show that $\partial \boldsymbol{f}/\partial \boldsymbol{p} = B \cdot \partial\boldsymbol{Q}/\partial\boldsymbol{p}$, which is readily computable, is an inconsistent set of equations for $\boldsymbol{f}(\boldsymbol{p})$.

Path-conservative schemes have been observed empirically (e.g.~\cite{dumbser2011simple, dumbser2016new}) to yield unique weak solutions provided that the non-conservative terms in Eq.~\eqref{eq:nonconsform} are zero across all genuinely nonlinear fields. Genuinely nonlinear fields are eigenvectors $\boldsymbol{K}_i$ of the Jacobian $\partial \boldsymbol{f}/\partial \boldsymbol{Q} + B$ such that $\vec{\nabla}_{\boldsymbol{Q}} \lambda_i(\boldsymbol{Q}) \cdot \boldsymbol{K}_i(\boldsymbol{Q}) \neq 0 \; \forall \; \boldsymbol{Q}$, where $\lambda_i$ is the corresponding eigenvalue and $\vec{\nabla}_{\boldsymbol{Q}}$ is a gradient with respect to the variables $\boldsymbol{Q}$~\cite{toro2013riemann}. The eigenvectors and eigenvalues in the 1-dimensional case in flat spacetime are
\begin{eqnarray}
\lambda_0 &=& v \nonumber\\
\lambda_{\pm} &=& \frac{v(1-\epsilon\Gamma(\Gamma-2)) \pm \rho_0^{-(2+\Gamma)}\sqrt{\epsilon h (\Gamma-1)\Gamma \rho_0^{4+2\Gamma} (1-v^2)^2} }{1+\epsilon\Gamma (1-(\Gamma-1) v^2)} \\
\boldsymbol{K}_0 &=& \left( \frac{\rho_0^{\Gamma-1}}{\epsilon(1-\Gamma)}, \frac{\Gamma \rho_0^{\Gamma-2} v}{\Gamma-1}, 1 \right)^{\top} \nonumber\\
\boldsymbol{K}_{\pm} &=& \left( \frac{\rho_0^{\Gamma-1}}{\epsilon}, \frac{\Gamma(\Gamma-1)h \rho_0^{2+\Gamma} \mp v\sqrt{ \epsilon h \Gamma(\Gamma-1) \rho_0^{4+2\Gamma} }}{-\epsilon \Gamma (\Gamma-1) \rho_0^4 v \pm \rho_0^{2-\Gamma}\sqrt{\epsilon h \Gamma (\Gamma-1) \rho_0^{4+2\Gamma}} } , 1 \right)^\top
\end{eqnarray}
The genuinely nonlinear fields are $\vec{K}_{\pm}$, and the non-conservative terms in Eq.~\eqref{eq:pcollected} are indeed zero across them (i.e. $B \cdot \vec{K}_{\pm} = 0$).

Despite this, we will find the weak solutions are unphysical. This can plausibly be blamed on the fact that the evolution variables are not the rest mass density, momentum density, and energy density~\cite{dumbserprivate,toro2013riemann}, and thus the implied jump conditions across discontinuities are physically incorrect. However, this explanation is not as straightforward as it would be for a conservative formulation, and so strictly speaking remains a speculation. Note that, in this work, we have not excluded the possibility of obtaining physically correct weak solutions using the path through state space that is consistent with the viscosity solutions of the system, since we use only the linear path presented in~\cite{dumbser2016new}.
%
%
\section{Numerical Results}\label{sec:results}
In this section we present shock tube solutions for the non-conservative Hamiltonian formulation of baroclinic fluid dynamics described by Eqs.~\eqref{eq:Dcollected}-\eqref{eq:DeScollected} and the ideal gas equation of state $P=\rho_0\epsilon(\Gamma-1)$. We use $\Gamma=2$. The domain length is $L=10$ with variables copied for the boundary conditions on either side, in flat spacetime in Cartesian coordinates. We use a Courant-Friedrichs-Lewy factor $\mathrm{cfl}=0.25$ unless otherwise specified, so that the time step and grid spacing are related by $\Delta t = \mathrm{cfl}\times \Delta x$. We use the path-conservative HLL scheme given explicitly in~\cite{dumbser2016new}. We obtain the exact Valencia solutions using~\cite{giacomazzo2006exact}. The solver is available to download from~\cite{brunosolver}. 

We will see that rarefaction fans are correctly captured, even superior to the Valencia formulation, as one would expect based on their being isentropic~\cite{toro2013riemann,marti1994analytical,lora2013exact} and the fact that our Hamiltonian formulation conserves the entropy density explicitly. Large discontinuities in $\rho_0$, without discontinuities in $\epsilon$, also appear to be very well-tolerated, although small deviations are notable. Other types of discontinuities yield clearly unphysical solutions. We will also compare with numerical solutions obtained using the Valencia formulation.

\subsection{Relativistic shock tube}
\begin{figure}
\centering
\includegraphics[width=1.0\textwidth]{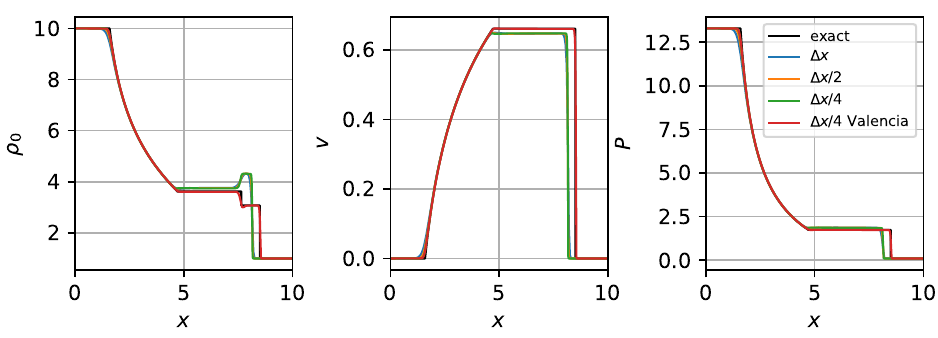}

\caption{Shock tube initial data $(\rho_{0,\mathrm{L}},v_{\mathrm{L}},P_{\mathrm{L}}) = (10,0,13.3)$, $(\rho_{0,\mathrm{R}},v_{\mathrm{R}},P_{\mathrm{R}}) = (1,0,0.1)$ at $t=4$ with fiducial resolution $\Delta x = 10/200$. For comparison, the numerical solution in the Valencia formulation is also displayed with resolution $\Delta x/4$.} \label{fig:reltube}
\end{figure}

This test is shown in Fig.~\ref{fig:reltube}. The left and right initial conditions we use for this evolution are $(\rho_{0,\mathrm{L}},v_{\mathrm{L}},P_{\mathrm{L}}) = (10,0,13.3)$, $(\rho_{0,\mathrm{R}},v_{\mathrm{R}},P_{\mathrm{R}}) = (1,0,0.1)$, respectively. The numerical solution in the Hamiltonian formulation is well-resolved, resulting in the three resolutions being difficult to distinguish. The contact discontinuity is not captured, since the pressure exhibits a jump there, whereas the only discontinuity should be in the rest mass density. The intermediate states to the left and right of the contact discontinuity at $x\sim 7.5$ are also unphysical, as well as the shockwave speed.

\subsection{Density discontinuity}
\begin{figure}
\centering
\includegraphics[width=1.0\textwidth]{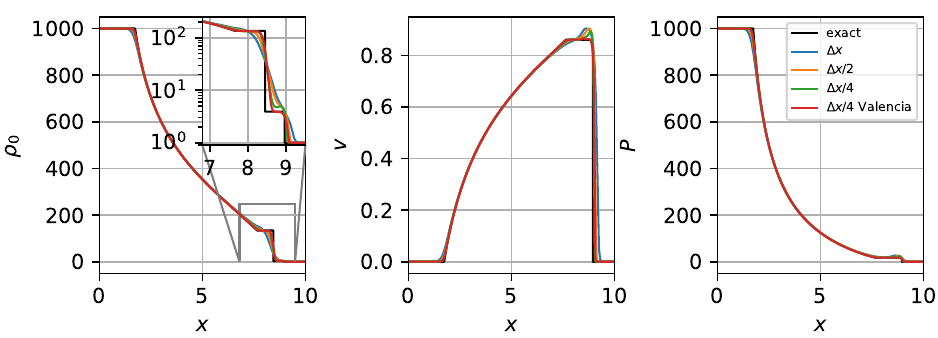}

\caption{Shock tube initial data $(\rho_{0,\mathrm{L}},v_{\mathrm{L}},\epsilon_{\mathrm{L}}) = (1000,0,1)$, $(\rho_{0,\mathrm{R}},v_{\mathrm{R}},\epsilon_{\mathrm{R}}) = (1,0,1)$ at $t=4$ with fiducial resolution $\Delta x = 10/200$. For comparison, the numerical solution in the Valencia formulation is also displayed, using the resolution $\Delta x/4$.} \label{fig:denstube}
\end{figure}

This test is shown in Fig.~\ref{fig:denstube}. The left and right initial conditions we use for this evolution are $(\rho_{0,\mathrm{L}},v_{\mathrm{L}},\epsilon_{\mathrm{L}}) = (1000,0,1)$, $(\rho_{0,\mathrm{R}},v_{\mathrm{R}},\epsilon_{\mathrm{R}}) = (1,0,1)$, respectively. The numerical solution in the Hamiltonian formulation is again well-resolved, and captures the exact solution well despite the very large initial discontinuity. However, the inset reveals a slightly incorrect density on the right side of the contact discontinuity.

\subsection{Specific internal energy discontinuity}
\begin{figure}
\centering
\includegraphics[width=1.0\textwidth]{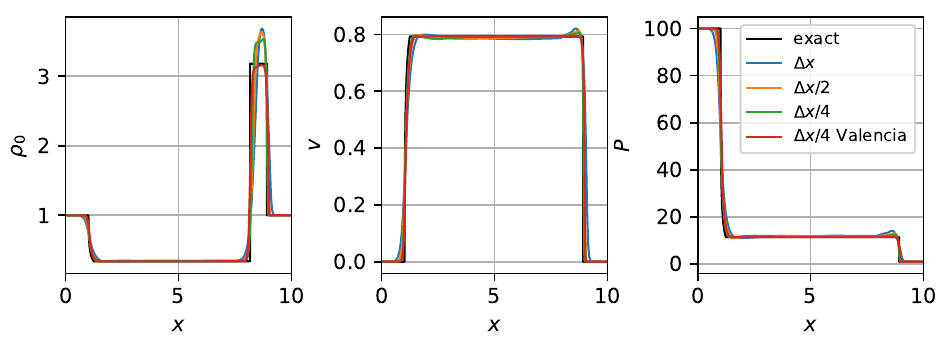}

\caption{Shock tube initial data $(\rho_{0,\mathrm{L}},v_{\mathrm{L}},\epsilon_{\mathrm{L}}) = (1,0,100)$, $(\rho_{0,\mathrm{R}},v_{\mathrm{R}},\epsilon_{\mathrm{R}}) = (1,0,1)$ at $t=4$ with fiducial resolution $\Delta x = 10/200$. For comparison, the numerical solution in the Valencia formulation is also displayed, using the resolution $\Delta x/4$.} \label{fig:epstube}
\end{figure}

This test is shown in Fig.~\ref{fig:epstube}. The left and right initial conditions we use for this evolution are $(\rho_{0,\mathrm{L}},v_{\mathrm{L}},\epsilon_{\mathrm{L}}) = (1,0,100)$, $(\rho_{0,\mathrm{R}},v_{\mathrm{R}},\epsilon_{\mathrm{R}}) = (1,0,1)$, respectively. The intermediate state (corresponding to the ``tower'' feature in the density) is unphysical.

\subsection{Velocity discontinuity}
\begin{figure}
\centering
\includegraphics[width=1.0\textwidth]{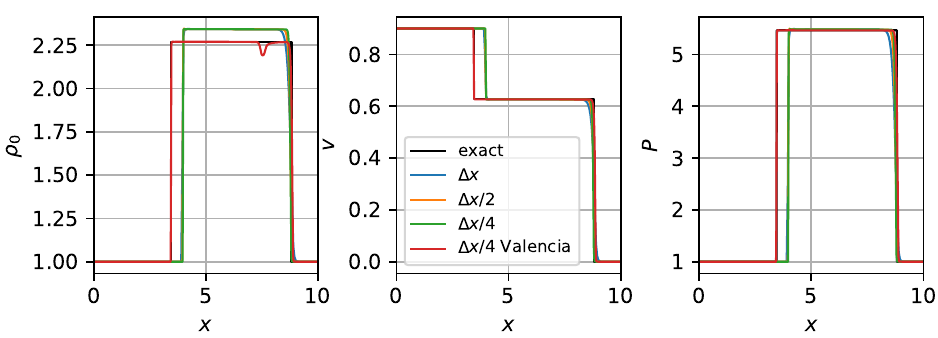}

\caption{Shock tube initial data $(\rho_{0,\mathrm{L}},v_{\mathrm{L}},\epsilon_{\mathrm{L}}) = (1,0.9,1)$, $(\rho_{0,\mathrm{R}},v_{\mathrm{R}},\epsilon_{\mathrm{R}}) = (1,0,1)$ at $t=4$ with fiducial resolution $\Delta x = 10/200$. For comparison, the numerical solution in the Valencia formulation is also displayed, using the resolution $\Delta x/4$.} \label{fig:veltube1}
\end{figure}

We consider two cases with the initial velocity in the left state being rightwards and leftwards. The first case is shown in Fig.~\ref{fig:veltube1}, whereby the left and right initial conditions we use for the rightwards evolution are $(\rho_{0,\mathrm{L}},v_{\mathrm{L}},\epsilon_{\mathrm{L}}) = (1,0.9,1)$, $(\rho_{0,\mathrm{R}},v_{\mathrm{R}},\epsilon_{\mathrm{R}}) = (1,0,1)$, respectively. The solution consists of a double shockwave, and the shockwave speeds and intermediate fluid state are seen to be unphysical. Resolving the intermediate state appears to be challenging for the Valencia scheme, where a dip in the density is exhibited near $x\sim 7.5$, which shrinks with increasing resolution.

\begin{figure}
\centering
\includegraphics[width=1.0\textwidth]{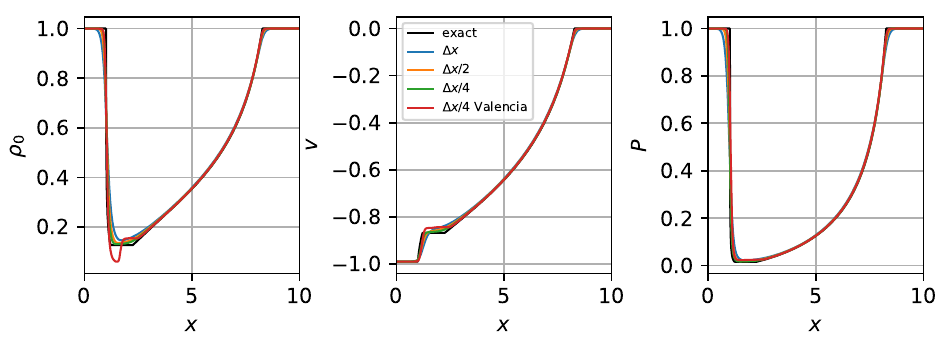}

\caption{Shock tube initial data $(\rho_{0,\mathrm{L}},v_{\mathrm{L}},\epsilon_{\mathrm{L}}) = (1,-0.99,1)$, $(\rho_{0,\mathrm{R}},v_{\mathrm{R}},\epsilon_{\mathrm{R}}) = (1,0,1)$ at $t=4$ with fiducial resolution $\Delta x = 10/200$. For comparison, the numerical solution in the Valencia formulation is displayed as well, using the resolution $\Delta x/4$.} \label{fig:veltube2}
\end{figure}

For the leftwards evolution displayed in Fig.~\ref{fig:veltube2}, we use $(\rho_{0,\mathrm{L}},v_{\mathrm{L}},\epsilon_{\mathrm{L}}) = (1,-0.99,1)$, $(\rho_{0,\mathrm{R}},v_{\mathrm{R}},\epsilon_{\mathrm{R}}) = (1,0,1)$, respectively. The solution consists of a double rarefaction, and it is captured well in the Hamiltonian scheme. Rarefaction fans are captured well in a variety of formulations because the solution is continuous and isentropic across the fans~\cite{toro2013riemann,marti1994analytical,lora2013exact}. But the Hamiltonian formulation we are using captures them particularly well, likely due to the fact that we are evolving the entropy density explicitly. The numerical Valencia solution actually performs worse, exhibiting a dip feature in the density near $x\sim 2.5$. This observation suggests that it may be beneficial in numerical applications to hybridize a numerical scheme to use Valencia near discontinuities, but evolve the entropy density otherwise.

%
%

\section{Conclusions} \label{sec:conc}
Hamiltonian formulations of fluid dynamics potentially offer unique advantages in numerical relativity, most of which are expected to be utilizable during the barotropic inspiral phase of a gravitational wave-driven binary coalescence involving at least one material body. The merger phase of the binary requires a baroclinic description. There are no advantages of using a Hamiltonian formulation over standard formulations that are currently apparent (although see~\cite{markakis2017conservation}).

Nonetheless, the baroclinic case is worth consideration in order to obtain a better understanding of the limitations of Hamiltonian formulations in future applications. This work is part of a series of papers exploring the viability of such formulations in practice. We have considered a Hamiltonian formulation of baroclinic fluid dynamics, showing that it yields unphysical weak solutions under a path-conservative approximate Riemann solver scheme. Thus, in future implementations of Hamiltonian formulations in numerical relativity, an explicitly barotropic formulation is advised during the inspiral phase, and then switching to a standard baroclinic formulation prior to merger will be necessary. However, note that we have not excluded the possibility of obtaining physically correct weak solutions using the path through state space that is consistent with the viscosity solutions of the system, since we use only the linear path presented in~\cite{dumbser2016new}.

In the appendix we also point out that the barotropic Hamiltonian formulation admits shockwaves at fluid-vacuum interfaces, which may be related to the numerical instabilities observed at the stellar surface in~\cite{westernacher2019hamilton}. Those instabilities were dealt with via a hybrid Hamiltonian-Valencia scheme.
%
%
\acknowledgements
We thank Charalampos Markakis, Vasileios Paschalidis, Eleuterio Toro, and Michael Dumbser for discussions. We give a special thanks to both Michael Dumbser and Dinshaw Balsara for providing code samples for the path-conservative HLL and HLLEM solvers in~\cite{dumbser2016new}, which greatly facilitated this work. This research is supported by National Science Foundation Grant No. PHY-1912619 at the University of Arizona.

\appendix

\section{Barotropic vacuum Riemann problem} \label{app:vac}
It is well-known that the Newtonian vacuum Riemann problem results in a pure rarefaction adjacent to the vacuum (see e.g.~\cite{munz1994tracking,munz1994numerical,toro2013riemann}). We can readily deduce the same conclusion in the special relativistic case using the jump conditions for the barotropic problem coming from conservation of rest mass and momentum,
\begin{eqnarray}
\left( \rho_L W_L - \rho_R W_R\right) S &=& \rho_L W_L v_L - \rho_R W_R v_R \\
\left( \rho_L h_L W_L^2 v_L - \rho_R h_R W_R^2 v_R\right) S &=& \rho_L h_L W_L^2 v_L^2 - P_L - \rho_R h_R W_R^2 v_R + p_R,
\end{eqnarray}
where $S$ is the shock speed, and the $_L$ and $_R$ subscripts denote the states to the left and right of the potential discontinuity. Setting the right state equal to vacuum, $\rho_R=P_R=v_R=0$, which implies $h_R=W_R=1$, the jump conditions reduce to
\begin{eqnarray}
S &=& v_L \\
0 &=& P_L.
\end{eqnarray}
Since the pressure in the fluid state vanishes, the pressure is therefore required to be continuous at the fluid vacuum interface. If the density must vanish along with the pressure (as per e.g.~$P=\kappa \rho_0^\Gamma$), then the fluid-vacuum interface cannot support a discontinuity in the density either. Shockwaves are therefore not supported at fluid-vacuum interfaces. A rarefaction is the only other possible elementary wave in the barotropic case (and this is still true in the baroclinic case~\cite{munz1994tracking,munz1994numerical,toro2013riemann}).

In the Hamiltonian formulation, the solution structure is different. Although all formulations agree in the rarefaction fan~\cite{toro2013riemann}, they do not necessarily agree on whether or where the rarefaction fan terminates. Consider the jump conditions in the Hamiltonian formulation coming from conservation of rest mass and the Hamiltonian Euler equation,
\begin{eqnarray}
\left( \rho_L W_L - \rho_R W_R\right) S &=& \rho_L W_L v_L - \rho_R W_R v_R \\
\left( h_L W_L v_L - h_R W_R v_R \right) S &=& h_L W_L - h_R W_R.
\end{eqnarray}
Setting the right state equal to vacuum yields
\begin{eqnarray}
S &=& v_L \\
h_L W_L v_L^2 &=& h_L W_L - 1,
\end{eqnarray}
or written another way,
\begin{eqnarray}
S &=& \mathrm{sign}(v_L) \sqrt{\frac{h_L^2 - 1}{h_L^2}} \\
h_L &=& W_L. 
\end{eqnarray}
We see that the specific enthalpy can take on values greater than 1, implying a positive rest mass density. Rest mass discontinuities are therefore supported as shockwaves at fluid-vacuum interfaces in the Hamiltonian formulation. This may be related to the numerical instabilities found at the stellar surface in~\cite{westernacher2019hamilton} when using a barotropic Hamiltonian formulation there.

\section{Convergence tests} \label{app:conv}

To validate our numerical implementations, we performed independent residual tests on smooth numerical solutions. This means plugging the numerical solutions into the fluid equations in a separately coded script, and observing the rate of decrease of the residual as resolution is increased. An independent residual test is a form of analytic convergence test not requiring specific knowledge of an exact solution, other than the fact that the solution must obey the equations of motion. 

\begin{figure}
\centering
\includegraphics[width=0.49\textwidth]{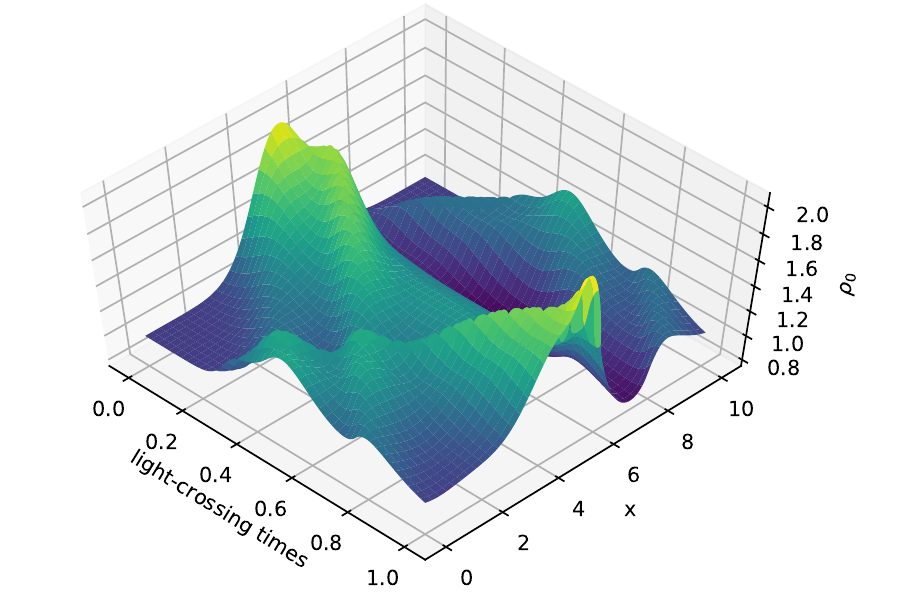}

\caption{The evolution of the rest mass density $\rho_0$ for the initial conditions~\eqref{eq:convic}, showing a highly dynamical solution.} \label{fig:evoconv}
\end{figure}

We use periodic boundary conditions on a domain length $L=10$ and a time step limitation $\mathrm{cfl}=0.25$. The initial conditions we use for this test are
\begin{eqnarray}
\rho_o &=& 1 + \exp{\left\lbrace \frac{-(x-0.5\times L)^2}{2}\right\rbrace } \nonumber\\
\epsilon &=& 1 + \exp{\left\lbrace\frac{-(x-0.7\times L)^2}{2\times 0.5^2}\right\rbrace} \label{eq:convic}\\
v &=& 0.2\times \left(x-0.5\times L\right) \exp{\left\lbrace \frac{-(x-0.5\times L)^2}{2\times 0.75^2} \right\rbrace},\nonumber
\end{eqnarray}
and the subsequent evolution of the rest mass density is depicted in Fig.~\ref{fig:evoconv}. This depiction is intended to convey that this is a very dynamical evolution, and is therefore a non-trivial test of the numerical implementation. We evolve until 1 light-crossing time.

For two residuals $R_{\Delta x}(x,t)$ and $R_{\Delta x/2}(x,t)$ obtained at resolutions $\Delta x$ and $\Delta x/2$, respectively, the global convergence factor is defined as
\begin{eqnarray}
(\log{2})^{-1} \log{\left\lbrace \frac{\mathcal{L}_2(R_{\Delta x})}{\mathcal{L}_2(R_{\Delta x/2})}\right\rbrace},
\end{eqnarray}
where we subjected the residuals to the $\mathcal{L}_2$-norm operator over space. The global convergence factor is a function of time, and its value is the convergence order measured in a spatially integrated sense.

\begin{figure}
\centering
\includegraphics[width=0.49\textwidth]{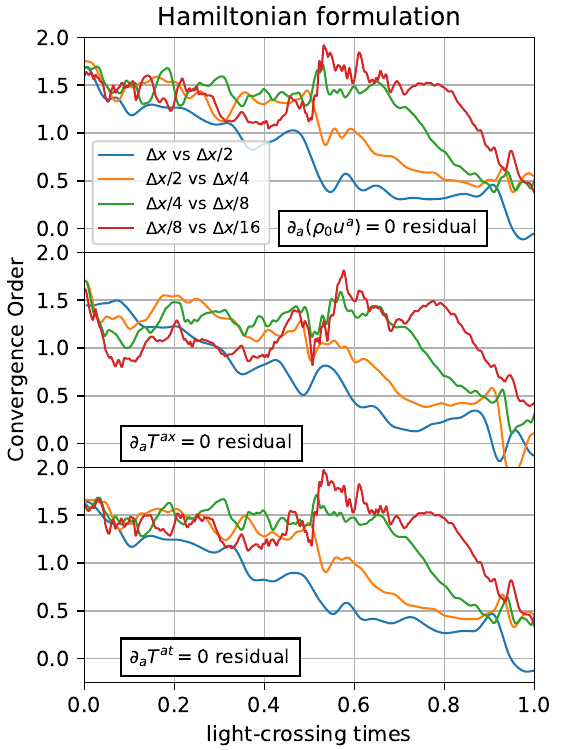}
\includegraphics[width=0.49\textwidth]{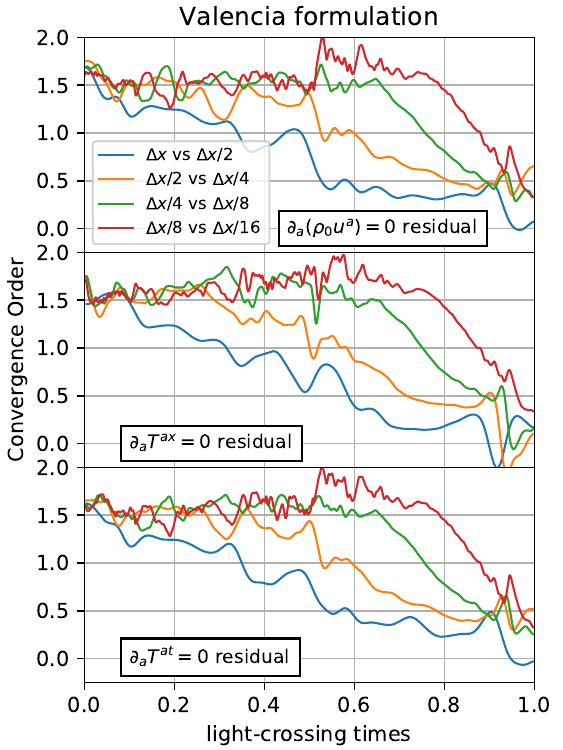}

\caption{Global convergence orders in the Hamiltonian (left) and Valencia (right) formulations for the initial conditions~\eqref{eq:convic}. The fiducial resolution is $\Delta x = 10/50$, and the minmod slope limiter is deployed in the numerical schemes. These plots are smoothed with a Gaussian kernel of width equal to 3 time steps for each resolution.} \label{fig:globconv}
\end{figure}

\begin{figure}
\centering
\includegraphics[width=0.49\textwidth]{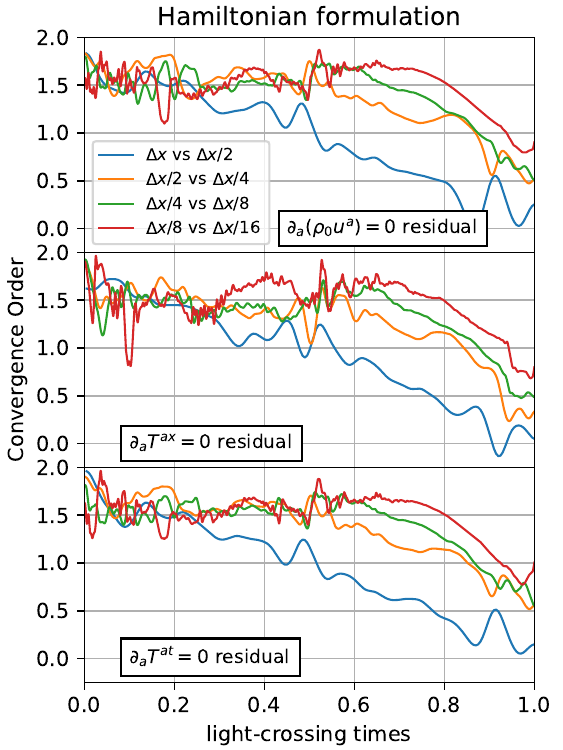}

\caption{Global convergence orders in the Hamiltonian formulation for the initial conditions~\eqref{eq:convic}. The fiducial resolution is $\Delta x = 10/50$, and the ``monotonized central'' (MC) slope limiter is deployed in the numerical scheme. These plots are smoothed with a Gaussian kernel of width equal to 3 time steps for each resolution.} \label{fig:globconvMC}
\end{figure}

In Fig.~\ref{fig:globconv} we display global convergence tests for the residuals of the fluid equations for both the Hamiltonian and Valencia formulations. The fiducial resolution is $\Delta x = 10/50$. The slope limiter deployed is the minmod type~\cite{toro2013riemann}, and in the Hamiltonian formulation the slope limiter is also applied to the non-conservative product ($B(\boldsymbol{Q})\partial_i \boldsymbol{Q}$ term in Eq.~\eqref{eq:nonconsform}), as per the numerical scheme in~\cite{dumbser2016new}. Both the Hamiltonian and Valencia formulations yield the expected $\sim$1.5th order global convergence for a nominally 2nd order finite volume scheme. However, the Euler equation residual in the Hamiltonian formulation suffers a somewhat diminished performance of 1st order during the first half of the evolution (Fig.~\ref{fig:globconv}, left middle panel). Based on our investigations, this diminished performance is due solely to the minmod-limited finite difference operator applied to the non-conservative term ($B(\boldsymbol{Q})\partial_i \boldsymbol{Q}$ term in Eq.~\eqref{eq:nonconsform}) in the Hamiltonian Euler equation, as per~\cite{dumbser2016new}. Using instead a ``monotonized central'' limiter (MC limiter)~\cite{toro2013riemann} yields an improved convergence order of $\sim$1.5 (see Fig.~\ref{fig:globconvMC}). However, we find the MC limiter performs more poorly on shockwave solutions in comparison to the minmod limiter, therefore the shock tube evolutions we present in this work always use the minmod limiter.

As time proceeds, numerical truncation error builds up in the solutions. This manifests in the decreasing convergence order at later times. As resolution is increased, the expected $\sim$1.5th order convergence is maintained for longer durations of time.

\begin{figure}
\centering
\includegraphics[width=0.49\textwidth]{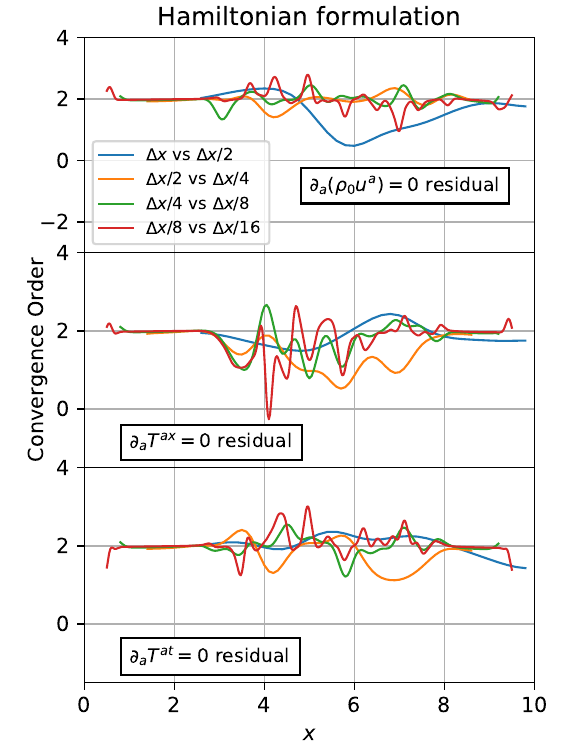}
\includegraphics[width=0.49\textwidth]{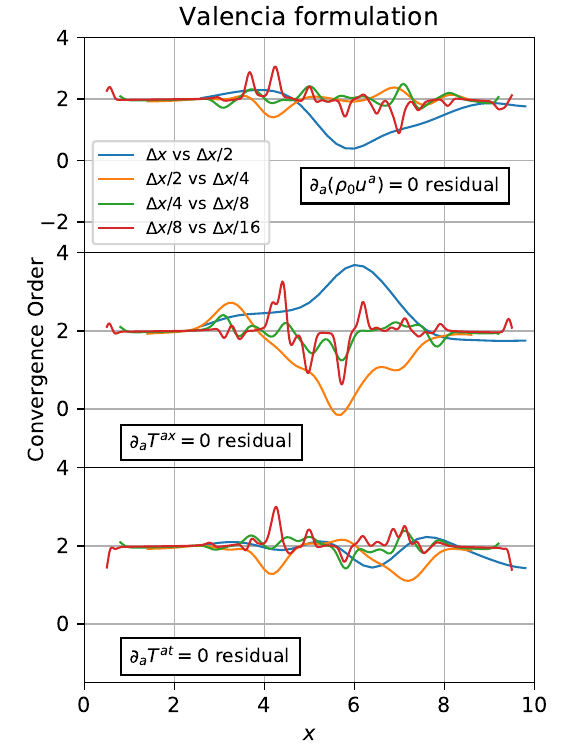}

\caption{Local convergence orders at $t=\Delta t$ in the Hamiltonian (left) and Valencia (right) formulations for the initial conditions~\eqref{eq:convic}. The fiducial resolution is $\Delta x = 10/50$, and the minmod slope limiter is deployed in the numerical schemes. The fiducial resolution is evolved by 1 time step, whereas the resolution $\Delta x/2$ is evolved by 2 time steps, etc.~such that all the residuals are compared at the same physical time. These plots are smoothed with a Gaussian kernel of width equal to 3 grid separations for each resolution.} \label{fig:locconv}
\end{figure}

We also display the convergence order for the spatially local residual in Fig.~\ref{fig:locconv}. This is defined as
\begin{eqnarray}
(\log{2})^{-1} \log{\left\lbrace \frac{\vert R_{\Delta x}(t=\Delta t,x) \vert}{\vert R_{\Delta x/2}(t=\Delta t,x) \vert}\right\rbrace},
\end{eqnarray}
where now we evaluate the residual at the physical time corresponding to the first time step in the fiducial resolution $t=\Delta t$, and take the absolute value rather than the $\mathcal{L}_2$-norm. The result is a function of spatial position, and indicates the instantaneous local convergence order. At such an early time, 2nd order convergence is obtained as expected.

\bibliography{fluidbib}

\end{document}